# DERIVATION OF EQUATIONS FOR SCALAR AND FERMION FIELDS USING PROPERTIES OF DISPERSION-CODISPERSION OPERATORS

RAOELINA ANDRIAMBOLOLONA[1], Ravo Tokiniaina RANAIVOSON [2], HANITRIARIVO Rakotoson[3], Victor HARISON[4]

*raoelinasp@yahoo.fr*[1]; *jacquelineraoelina@hotmail.com*[1]
*tokhiniaina@gmail.com*[2]; *infotsara@gmail.com*[3]; *Vharison@inscae.mg* [4]

*Theoretical Physics Department*[1,2,3]

*Institut National des Sciences et Techniques Nucléaires (INSTN- Madagascar)*

BP 4279 101, Antananarivo -Madagascar, *instn@moov.mg*

*Institut National des Sciences Comptables et de l'Administration d'Entreprises*[4] *(INSCAE),* Antananarivo-Madagascar

**Abstract:** We establish equations for scalar and fermion fields using results obtained from a study on a phase space representation of quantum theory that we have performed in a previous work. Our approaches are similar to the historical ones to obtain Klein-Gordon and Dirac equations but the main difference is that ours are based on the use of properties of operators called dispersion-codispersion operators. We begin with a brief recall about the dispersion-codispersion operators. Then, introducing a mass operator with its canonical conjugate coordinate and applying rules of quantization, based on the use of dispersion - codispersion operators , we deduce a second order differential operator relation from the relativistic expression relying energy, momentum and mass. Using Dirac matrices, we derive from this second order differential operator relation a first order one. The application of the second order differential operator relation on a scalar function gives the equation for the scalar field and the use of the first order differential operator relation leads to the equation for fermion field.

## 1-INTRODUCTION

Using results obtained in our previous work [1] concerning a study on a phase space representation of quantum theory, our aim in this paper is to obtain equations for scalar and fermion fields. Our approaches have some similarities with the historical ones used for the obtention of the Klein-Gordon and Dirac equations [2],[3],[4],[5] but there are some differences. In fact, our approaches are based on the use of the properties of operators called dispersion-codispersion operators defined in [1].

In the reference [1], statistic-probability theory and linear algebra are both used. We reserve the word "covariance" for its meaning in linear algebra [8] (variance and covariance of a tensor) and the words "dispersion-codispersion" for the statistical variance-covariance. It is to be noted also that operators are denoted with bold letters. The natural unit system for quantum field theory ($\hbar = 1, c = 1$) is used.

It is well known that the standard equations in field theory for scalar and fermion fields are respectively Klein-Gordon and Dirac equations. These field equations are obtained by combining quantum theory and special relativity [3], [4], [5], [6], [7]. Let us consider the relativistic relation between energy, momentum and mass

$$(E)^2 - (\vec{p})^2 - m^2 = 0 \iff g^{\mu\nu}p_\mu p_\nu - m^2 = 0 \quad (\mu = 0, 1, 2, 3) \quad (1.1)$$

$$g^{00} = 1 \quad g^{jj} = -1 \text{ for } j = 1, 2, 3 \quad g^{\mu\nu} = 0 \text{ for } \mu \neq \nu$$

By using the quantization rules which consists to replace the components of the energy-momentum quadricovector by the corresponding operators

$$p_\mu \to i\frac{\partial}{\partial x^\mu} \tag{1.2}$$

we obtain a second order differential operator relation

$$g^{\mu\nu}\frac{\partial}{\partial x^\mu}\frac{\partial}{\partial x^\nu} + m^2 = 0 \Leftrightarrow \frac{\partial^2}{(\partial x^0)^2} - \frac{\partial^2}{(\partial x^1)^2} - \frac{\partial^2}{(\partial x^2)^2} - \frac{\partial^2}{(\partial x^3)^2} + m^2 = 0 \tag{1.3}$$

On one hand, the application of this operator relation on a scalar function $\phi$ gives the Klein-Gordon equation

$$(g^{\mu\nu}\frac{\partial}{\partial x^\mu}\frac{\partial}{\partial x^\nu} + m^2)\phi = 0 \tag{1.4}$$

On the other hand, by using Dirac matrices $\gamma^\mu$ which verify the anticommutation properties

$$\gamma^\mu\gamma^\nu + \gamma^\nu\gamma^\mu = 2g^{\mu\nu}I_4 \tag{1.5}$$

, in which $I_4$ is the $4 \times 4$ identity matrix, we can deduce from the relation (1.3) a first order differential operator relation

$$\left(i\gamma^\mu\frac{\partial}{\partial x^\mu} - m\right) = 0 \tag{1.6}$$

In fact, we have the factorization relation

$$g^{\mu\nu}\frac{\partial}{\partial x^\mu}\frac{\partial}{\partial x^\nu} + m^2 = -\left(i\gamma^\mu\frac{\partial}{\partial x^\mu} + m\right)\left(i\gamma^\nu\frac{\partial}{\partial x^\nu} - m\right) \tag{1.7}$$

The application of the operator relation (1.6) on a spinor function $\psi$ leads to the Dirac equation

$$\left(i\gamma^\mu\frac{\partial}{\partial x^\mu} - m\right)\psi = 0 \tag{1.8}$$

In the beginning, the Klein-Gordon and Dirac equations were expected to be just the relativistic equivalent of the Schrödinger equation. But it was later seen that the functions which appear in these equations are not to be interpreted as wave functions but as fields and the equations themselves have their right place and interpretations only in the framework of field theory [3], [4], [5], [6], [7].

As said, we establish in this work other equations for scalar and fermion fields which may be considered as similar to the Klein-Gordon and Dirac equations. These equations are given in the relations (3.13) and (4.20).

## 2-RECALL ABOUT DISPERSION-CODISPERSION OPERATORS

In our work [1], we have introduced operators called dispersion operators. For the one dimensional case, the expression of a momentum dispersion operator is

$$\Sigma = \frac{1}{2}\left[\frac{(\boldsymbol{p}-P)^2}{(\Delta p)^2} + \frac{(\boldsymbol{x}-X)^2}{(\Delta x)^2}\right](\Delta p)^2 = \frac{1}{2}\left[(\boldsymbol{p}-P)^2 + \frac{(\Delta p)^2}{(\Delta x)^2}(\boldsymbol{x}-X)^2\right] \quad (2.1)$$

the operator $\Sigma$ admits as eigenstates the states denoted $|n,X,P,\Delta p\rangle$ which are states whose corresponding wave functions in the coordinate and momentum representation are harmonic Gaussian functions $\varphi_n(x,X,P,\Delta p)$ and their Fourier transforms [1]

$$\langle x|n,X,P,\Delta p\rangle = \varphi_n(x,X,P,\Delta p) = \frac{H_n(\frac{x-X}{\sqrt{2}\Delta x})}{\sqrt{2^n n! \sqrt{2\pi}\Delta x}} e^{-(\frac{x-X}{2\Delta x})^2 + iPx} \quad (2.2)$$

$$\langle p|n,X,P,\Delta p\rangle = \tilde{\varphi}_n(p,X,P,\Delta p) = \frac{1}{\sqrt{2\pi}}\int \varphi_n(x,X,P,\Delta p) e^{-ipx} dx \quad (2.3)$$

in these expressions,

- ➤ $\boldsymbol{x}$ and $\boldsymbol{p}$ are respectively the position and momentum operators. In the coordinate representation, we have

$$\boldsymbol{x} = x \qquad \boldsymbol{p} = -i\frac{d}{dx} \quad (2.4)$$

and in the momentum representation

$$\boldsymbol{x} = i\frac{d}{dp} \qquad \boldsymbol{p} = p \quad (2.5)$$

- ➤ $X$ and $P$ are respectively the coordinate and momentum mean values corresponding to the states $|n,X,P,\Delta p\rangle$

$$X = \langle n,X,P,\Delta p|\boldsymbol{x}|n,X,P,\Delta p\rangle = \int x\,|\varphi_n(x,X,P,\Delta p)|^2 dx \quad (2.6)$$

$$P = \langle n,X,P,\Delta p|\boldsymbol{p}|n,X,P,\Delta p\rangle = \int p\,|\tilde{\varphi}_n(p,X,P,\Delta p)|^2 dp \quad (2.7)$$

- ➤ $(\Delta x)^2$ and $(\Delta p)^2$ are respectively the coordinate and momentum dispersions (commonly called statistical variances) corresponding to the states $|0,X,P,\Delta p\rangle$

$$(\Delta x)^2 = \langle 0,X,P,\Delta p|(\boldsymbol{x}-X)^2|0,X,P,\Delta p\rangle = \int (x-X)^2\,|\varphi_0(x,X,P,\Delta p)|^2 dx \quad (2.8)$$

$$(\Delta p)^2 = \langle 0,X,P,\Delta p|(\boldsymbol{p}-P)^2|0,X,P,\Delta p\rangle = \int (p-P)^2\,|\tilde{\varphi}_0(p,X,P,\Delta p)|^2 dp \quad (2.9)$$

As $\tilde{\varphi}_0(p,X,P,\Delta p)$ is the Fourier transform of $\varphi_0(x,X,P,\Delta p)$, $\Delta x$ and $\Delta p$ are related by the relation:

$$\Delta x \Delta p = \frac{1}{2} \quad (2.10)$$

$(\Delta x)^2$ and $(\Delta p)^2$ are called respectively coordinate and momentum ground dispersions. For a state $|n, X, P, \Delta p\rangle$ the corresponding values of the coordinate and momentum dispersions are

$$(\Delta x_n)^2 = \langle n, X, P, \Delta p|(x - X)^2|n, X, P, \Delta p\rangle$$

$$= \int (x - X)^2 |\varphi_n(x, X, P, \Delta p)|^2 dx = (2n + 1)(\Delta x)^2 \qquad (2.11)$$

$$(\Delta p_n)^2 = \langle n, X, P, \Delta p|(p - P)^2|n, X, P, \Delta p\rangle$$

$$= \int (p - P)^2 |\tilde{\varphi}_n(p, X, P, \Delta p)|^2 dp = (2n + 1)(\Delta p)^2 \qquad (2.12)$$

It can be shown [1] that the eigenvalue of the momentum dispersion operator $\boldsymbol{\Sigma}$ (1.1) corresponding to an eigenstate $|n, X, P, \Delta p\rangle$ is equal to the value of momentum dispersion corresponding to this state

$$\boldsymbol{\Sigma}|n, X, P, \Delta p\rangle = (\Delta p_n)^2|n, X, P, \Delta p\rangle = (2n + 1)(\Delta p)^2|n, X, P, \Delta p\rangle \qquad (2.13)$$

From the momentum dispersion operator $\boldsymbol{\Sigma}$, we may define the momentum quadratic mean operator

$$\overline{p^2} = P^2 + \boldsymbol{\Sigma} \qquad (2.14)$$

$\overline{p^2}$ admits the same eigeinstates as $\boldsymbol{\Sigma}$

$$\overline{p^2}|n, X, P, \Delta p\rangle = (P^2 + \boldsymbol{\Sigma})|n, X, P, \Delta p\rangle = [P^2 + (2n + 1)(\Delta p)^2]|n, X, P, \Delta p\rangle \qquad (2.15)$$

For the case of quadridimensional relativistic theory, in Minkowski space, it has been shown in [1] that as generalization of the momentum dispersion $(\Delta p)^2$ we have to define momentum dispersion –codispersion tensor (equivalent to variance-covariance matrix in probability and statistics) $\mathcal{B}_{\mu\nu}$. If the variables are uncorrelated, the dispersion-codispersion tensor is diagonal:

$$\mathcal{B}_{\mu\mu} = (\Delta p_\mu)^2 = \frac{1}{2}\frac{\Delta p_\mu}{\Delta x^\mu} = \frac{1}{(2\Delta x^\mu)^2} \qquad \mathcal{B}_{\mu\nu} = 0 \; if \; \mu \neq \nu \; (\mu = 0, 1, 2, 3) \qquad (2.16)$$

If the variables are correlated, this tensor is not diagonal: $\mathcal{B}_{\mu\nu} \neq 0 \; if \; \mu \neq \nu$. Then as generalization of the momentum dispersion operator $\boldsymbol{\Sigma}$, it was also shown in [1] that we have to define a dispersion- codispersion tensor operator $\boldsymbol{\Sigma}_{\mu\nu}$

$$\boldsymbol{\Sigma}_{\mu\nu} = \frac{1}{2}(\boldsymbol{p}_\mu - P_\mu)(\boldsymbol{p}_\nu - P_\nu) + 2\mathcal{B}_{\mu\alpha}\mathcal{B}_{\nu\beta}(x^\alpha - X^\alpha)(x^\beta - X^\beta) \qquad (2.17)$$

➢ The $x^\mu$ and $p_\mu$ are respectively the operators corresponding to the coordinates and the components of the energy-momentum quadricovector. They obey the following commutation relations ($g_{\mu\nu}$ are the components of the metric tensor) :

$$[\boldsymbol{p}_\mu, \boldsymbol{x}_\nu]_- = ig_{\mu\nu} \qquad [\boldsymbol{p}_\mu, \boldsymbol{p}_\nu]_- = 0 \qquad [\boldsymbol{x}_\mu, \boldsymbol{x}_\nu]_- = 0 \qquad (2.18)$$

$$g_{00} = 1 \quad g_{kk} = -1 \text{ for } k = 1,2,3 \quad g_{\mu\nu} = 0 \text{ if } \mu \neq \nu \quad (2.19)$$

In the coordinate representation, we have

$$\boldsymbol{x}^\mu = x^\mu \quad \boldsymbol{p}_\mu = i\frac{\partial}{\partial x^\mu} \quad (2.20)$$

In the momentum representation, we have

$$\boldsymbol{x}^\mu = -i\frac{\partial}{\partial p_\mu} \quad \boldsymbol{p}_\mu = p_\mu \quad (2.21)$$

➢ The $X^\mu$ and the $P_\nu$ are respectively the mean values of the operators $\boldsymbol{x}^\mu$ and $\boldsymbol{p}_\mu$ when the state of a particle is an eigenstate $|n, X^\mu, P_\nu, \mathcal{B}_{\alpha\beta}\rangle$ of the dispersion- codispersion operator $\Sigma_{\mu\nu}$[1]

$$X^\mu = \langle n, X^\mu, P_\nu, \mathcal{B}_{\alpha\beta}|\boldsymbol{x}^\mu|n, X^\mu, P_\nu, \mathcal{B}_{\alpha\beta}\rangle \quad (2.22)$$

$$P_\nu = \langle n, X^\mu, P_\nu, \mathcal{B}_{\alpha\beta}|\boldsymbol{p}_\nu|n, X^\mu, P_\nu, \mathcal{B}_{\alpha\beta}\rangle \quad (2.23)$$

From the dispersion-codispersion operators $\Sigma_{\mu\nu}$, we can define quadratic means operators $\overline{(\boldsymbol{p}_\mu)^2}$:

$$\overline{(\boldsymbol{p}_\mu)^2} = (P_\mu)^2 + \Sigma_{\mu\mu} \quad (2.24)$$

## 3- EQUATION FOR SCALAR FIELD

Let us consider the relativistic relation between the energy-momentum and mass

$$g^{\mu\nu}p_\mu p_\nu = m^2 \Leftrightarrow (p_0)^2 - (p_1)^2 - (p_2)^2 - (p_3)^2 = m^2 \quad (3.1)$$

For the quantization of this relation in the framework of our approach, we adopt the following hypothesis:

***Hypothesis 1:*** There is a mass operator $\boldsymbol{m}$ and a canonical conjugate coordinate associated with $\boldsymbol{m}$, denoted $\boldsymbol{\tau}$, such we have the commutation relation

$$[\boldsymbol{m}, \boldsymbol{\tau}]_- = i \quad (3.2)$$

in the $\boldsymbol{\tau}$-representation

$$\boldsymbol{\tau} = \tau \quad \boldsymbol{m} = i\frac{\partial}{\partial \tau} \quad (3.3)$$

in the $\boldsymbol{m}$-representation

$$\boldsymbol{m} = m \quad \boldsymbol{\tau} = -i\frac{\partial}{\partial m} \quad (3.4)$$

Then we introduce the mass dispersion and mass quadratic mean operators

$$\overline{m^2} - M^2 = \frac{1}{2}\left[(\boldsymbol{m} - M)^2 + \frac{(\Delta m)^2}{(\Delta \tau)^2}(\boldsymbol{\tau} - \mathrm{T})^2\right] \tag{3.5}$$

$$\overline{m^2} = M^2 + \frac{1}{2}\left[(\boldsymbol{m} - M)^2 + \frac{(\Delta m)^2}{(\Delta \tau)^2}(\boldsymbol{\tau} - \mathrm{T})^2\right] \tag{3.6}$$

In these expressions, we assume that $\Delta m$ and $\Delta \tau$ are related by the relation

$$\Delta m \Delta \tau = \frac{1}{2} \Leftrightarrow \Delta \tau = \frac{1}{2\Delta m}$$

So, we have also for the operator $\overline{m^2}$

$$\overline{m^2} = M^2 + \frac{1}{2}(\boldsymbol{m} - M)^2 + 2(\Delta m)^4(\boldsymbol{\tau} - \mathrm{T})^2 \tag{3.7}$$

$M$ and $T$ are respectively the mean values of the operators $\boldsymbol{m}$ and $\boldsymbol{\tau}$ when the state is an eigenstate of the mass dispersion and mass quadratic mean operators.

***Hypothesis 2:*** For the quantization of the relation (3.1), the $(p_\mu)^2$ are replaced by the operators $\overline{(\boldsymbol{p_\mu})^2}$ and $m^2$ is replaced by $\overline{m^2}$. We then obtain the following operatorial relation

$$g^{\mu\nu}\Sigma_{\mu\nu} + g^{\mu\nu}P_\mu P_\nu - \overline{m^2} = 0 \tag{3.8}$$

We may suppose that between the mean values of the momentum components and the mass, we have the following relations

$$g^{\mu\nu}P_\mu P_\nu = M^2 \tag{3.9}$$

Then the relation (3.8) becomes

$$g^{\mu\nu}\Sigma_{\mu\nu} = \overline{m^2} - M^2 \tag{3.10}$$

The application of this operator relation on a scalar function lead to the an equation for the scalar field

$$\left[g^{\mu\nu}\Sigma_{\mu\nu} - (\overline{m^2} - M^2)\right]\phi = 0 \tag{3.11}$$

In coordinate representation, we have

$$\Sigma_{\mu\nu} = \frac{1}{2}\left(i\frac{\partial}{\partial x^\mu} - P_\mu\right)\left(i\frac{\partial}{\partial x^\nu} - P_\nu\right) + 2\mathcal{B}_{\mu\alpha}\mathcal{B}_{\nu\beta}(x^\alpha - X^\alpha)(x^\beta - X^\beta) \tag{3.12a}$$

$$\overline{m^2} = M^2 + \frac{1}{2}(\boldsymbol{m} - M)^2 + 2(\Delta m)^4(\boldsymbol{\tau} - \mathrm{T})^2 \tag{3.12b}$$

then the equation (3.11) becomes

$$\{g^{\mu\nu}\left[\left(i\frac{\partial}{\partial x^{\mu}}-P_{\mu}\right)\left(i\frac{\partial}{\partial x^{\nu}}-P_{\nu}\right)+4\mathcal{B}_{\mu\alpha}\mathcal{B}_{\nu\beta}(x^{\alpha}-X^{\alpha})(x^{\beta}-X^{\beta})\right]$$
$$-\left[\left(i\frac{\partial}{\partial\tau}-M\right)^{2}+4(\Delta m)^{4}(\tau-\mathrm{T})^{2}\right]\}\phi=0 \qquad (3.13)$$

we remark that according to the equation(3.13), the field $\phi$ is a function of the five variables $x^0, x^1, x^2, x^3$ and $\tau$.

## 4- EQUATION FOR FERMION FIELD

Let us consider the relation (3.10)

$$g^{\mu\nu}\Sigma_{\mu\nu}=\left(\overline{m^{2}}-M^{2}\right) \qquad (4.1)$$

If we look at the expressions (2.17) and (3.7) of $\Sigma_{\mu\nu}$ and $\overline{m^2}$, we remark that the relation (4.1) is a relation between second order differential operators. In this paragraph, our aim is to deduce from this relation a linear relation between first order differential operators with a view to obtain a first order differential equation for fermion field. As for the establishment of the Dirac equation, we consider the following "operator factorization":

$$g^{\mu\nu}\Sigma_{\mu\nu}-\left(\overline{m^{2}}-M^{2}\right)$$

$$=\frac{1}{2}g^{\mu\nu}[(p_{\mu}-P_{\mu})(p_{\nu}-P_{\nu})+4\mathcal{B}_{\mu\alpha}\mathcal{B}_{\nu\beta}(x^{\alpha}-X^{\alpha})(x^{\beta}-X^{\beta})]$$
$$-\frac{1}{2}[(m-M)^{2}+4(\Delta m)^{4}(\tau-\mathrm{T})^{2}] \qquad (4.2)$$

$$=\frac{1}{2}[\alpha^{\mu}(p_{\mu}-P_{\mu})+2\beta^{\mu}\mathcal{B}_{\mu\alpha}(x^{\alpha}-X^{\alpha})+\zeta(m-M)+2(\Delta m)^{2}\vartheta(\tau-\mathrm{T})]$$
$$\cdot[\alpha^{\nu}(p_{\nu}-P_{\nu})+2\beta^{\nu}\mathcal{B}_{\nu\beta}(x^{\beta}-X^{\beta})-\zeta(m-M)-2(\Delta m)^{2}\vartheta(\tau-\mathrm{T})]$$

$$=\frac{1}{2}\{\frac{\alpha^{\mu}\alpha^{\nu}+\alpha^{\nu}\alpha^{\mu}}{2}(p_{\mu}-P_{\mu})(p_{\nu}-P_{\nu})+2(\beta^{\mu}\beta^{\nu}+\beta^{\nu}\beta^{\mu})\mathcal{B}_{\mu\alpha}\mathcal{B}_{\nu\beta}(x^{\alpha}-X^{\alpha})(x^{\beta}-X^{\beta})$$
$$-\zeta^{2}(m-M)^{2}-4(\Delta m)^{4}\vartheta^{2}(\tau-\mathrm{T})^{2}+2(\alpha^{\mu}\beta^{\nu}+\beta^{\nu}\alpha^{\mu})\mathcal{B}_{\nu\beta}(p_{\mu}-P_{\mu})(x^{\beta}-X^{\beta})$$
$$+(\zeta\alpha^{\mu}-\alpha^{\mu}\zeta)(p_{\mu}-P_{\mu})(m-M)+2(\vartheta\alpha^{\mu}-\alpha^{\mu}\vartheta)(\Delta m)^{2}(p_{\mu}-P_{\mu})(\tau-\mathrm{T})$$
$$+2(\zeta\beta^{\mu}-\beta^{\mu}\zeta)\mathcal{B}_{\mu\alpha}(x^{\alpha}-X^{\alpha})(m-M)+4(\vartheta\beta^{\mu}-\beta^{\mu}\vartheta)\mathcal{B}_{\mu\alpha}(\Delta m)^{2}(x^{\alpha}-X^{\alpha})(\tau-\mathrm{T})$$
$$-2(\zeta\vartheta+\vartheta\zeta)(\Delta m)^{2}(m-M)(\tau-\mathrm{T})-2i\beta^{\mu}\alpha^{\nu}\mathcal{B}_{\mu\nu}+2i\vartheta\zeta(\Delta m)^{2}]\} \qquad (4.3)$$

In these relations, the coefficients $\alpha^{\mu},\beta^{\mu},\zeta$ and $\vartheta$ are considered as matrices which are to be determined.

By making identification between the relations (4.2) and (4.3), we can deduce the following relations:

$$\alpha^\mu \alpha^\nu + \alpha^\nu \alpha^\mu = 2g^{\mu\nu} \quad (4.4)$$

$$\beta^\mu \beta^\nu + \beta^\nu \beta^\mu = 2g^{\mu\nu} \quad (4.5)$$

$$\alpha^\mu \beta^\nu + \beta^\nu \alpha^\mu = 0 \quad (4.6)$$

$$\zeta^2 = 1 \quad (4.7)$$

$$\vartheta^2 = 1 \quad (4.8)$$

$$\zeta\vartheta + \zeta\vartheta = 0 \quad (4.9)$$

$$\zeta\alpha^\mu - \alpha^\mu\zeta = 0 \quad (4.10)$$

$$\vartheta\alpha^\mu - \alpha^\mu\vartheta = 0 \quad (4.11)$$

$$\zeta\beta^\mu - \beta^\mu\zeta = 0 \quad (4.12)$$

$$\vartheta\beta^\mu - \beta^\mu\vartheta = 0 \quad (4.13)$$

$$\beta^\mu \alpha^\nu \mathcal{B}_{\mu\nu} - \vartheta\zeta(\Delta m)^2 = 0 \quad (4.14)$$

We can show that a set of matrices which verify the relations (4.4) to (4.13) are:

$$\alpha^\mu = \gamma^\mu \otimes I_4 \otimes I_2 \quad (4.15)$$

$$\beta^\mu = \gamma^5 \otimes \gamma^\mu \otimes I_2 \quad (4.16)$$

$$\zeta = I_4 \otimes I_4 \otimes \begin{pmatrix} 1 & 0 \\ 0 & -1 \end{pmatrix} \quad (4.17)$$

$$\vartheta = I_4 \otimes I_4 \otimes \begin{pmatrix} 0 & 1 \\ 1 & 0 \end{pmatrix} \quad (4.18)$$

The relation (4.14) is a relation which relate $\mathcal{B}_{\mu\nu}$ and $(\Delta m)^2$.

Then, the equation for fermion field which can be deduced is

$$\left[\alpha^\mu(p_\mu - P_\mu) + 2\beta^\mu \mathcal{B}_{\mu\nu}(x^\nu - X^\nu) - \zeta(m - M) - 2(\Delta m)^2 \vartheta(\tau - \mathrm{T})\right]\psi = 0 \quad (4.19)$$

or

$$\left[\alpha^\mu(i\partial_\mu - P_\mu) + 2\beta^\mu \mathcal{B}_{\mu\nu}(x^\nu - X^\nu) - \zeta(i\frac{\partial}{\partial\tau} - M) - 2(\Delta m)^2 \vartheta(\tau - \mathrm{T})\right]\psi = 0 \quad (4.20)$$

According to the equation (4.20), the field $\psi$ is a function which depends on the five variables $x^0, x^1, x^2, x^3$ and $\tau$.

## 5-CONCLUSION

As it is shown by the equations (3.13) and (4.20), our approaches, which are based on the use of properties of dispersion-codispersion operators lead as expected to the obtention of equations for scalar and fermion fields. These equations have some similarities with the Klein-Gordon and Dirac equations but there are also some differences between them. As examples, unlike the case of the fields in the equations of Klein-Gordon and Dirac, which depend on the four variables $x^0, x^1, x^2, x^3$, the fields in our equations depend on five variables $x^0, x^1, x^2, x^3$ and $\tau$. Another remarkable difference is the explicit presence of the momentum dispersion-codispersion tensor $\mathcal{B}_{\mu\nu}$ and the mass dispersion $(\Delta m)^2$ in our equations. These presences are related to the fact that our approaches are based on the use of results from a phase space representation of quantum theory which takes into account the quantum uncertainty relation. Our approach may be then used in the formulation of a quantum field theory in phase space.